%% file: manuscript.tex
\documentclass[comma,a4paper]{article}
\input{preamble/preamble}

\begin{document}


\input{abstract}

\section{Introduction}
\input{intro}
\label{intro}
\section{The partial copula}
\label{partial_copula}
\input{partial_copula}
\section{Examples of partial copulas}
\label{partial_examples}
\input{partial_examples}

\input{fig_partial_frank}

\section{Properties of the partial copula}
\label{partial_cop_properties}
\input{partial_cop_properties}
\section{Properties of partial dependence measures}
\label{dependence_measures}
\input{dependence_measures}

\section{Concluding remarks}
\input{conclusion}

{\singlespacing
\bibliographystyle{model2-names}
\bibliography{manuscript}
}
\appendix
\section*{Appendix}
\renewcommand{\thesubsection}{\Alph{subsection}}
\input{proof_partial_frank}
\input{proof_l2distance}
\input{proof_cop_approx}
\input{proof_joint_seq}
\input{proof_not_archimedean}
\input{proof_cond_indep}
\input{proof_cond_corr}
\input{proof_kendall}

\end{document}

%% file: abstract.tex
\title{\vspace{-2cm}The partial copula: Properties and associated dependence measures
}
\author{Fabian Spanhel\thanks{Corresponding author.} \thanks{Department of Statistics, Ludwig-Maximilians-Universit{\"a}t M{\"u}nchen, Akademiestr. 1, 80799 Munich, Germany (email: \href{mailto:spanhel@stat.uni-muenchen.de}{spanhel@stat.uni-muenchen.de} and \href{mailto:malte.kurz@stat.uni-muenchen.de}{malte.kurz@stat.uni-muenchen.de}).}
\phantom{l}and
Malte S. Kurz\footnotemark[2]}


\maketitle

\begin{abstract}
The partial correlation coefficient is a commonly used measure to assess the conditional dependence between two random variables.
We provide a thorough explanation of the partial copula, which is a natural generalization of the partial correlation coefficient, and investigate several of its properties. 
In addition, properties of some associated partial dependence measures are examined.

\bigskip
\noindent \textbf{Keywords:}
Partial copula, Conditional copula, Partial correlation, Partial Spearman's rho, Partial Kendall's tau.
\end{abstract}

%% file: intro.tex
Studying the dependence between two random variables $Y_1$ and $Y_2$ conditional on a random vector $Z$ is an important topic in statistics. 
The partial correlation coefficient is often used to measure the conditional dependence between two random variables due to its simple computation and meaningful interpretation if the joint distribution of $(Y_{1:2},Z)$ is given by an elliptical distribution. 
However, outside the elliptical world, the interpretation of the partial correlation coefficient as a measure for conditional dependence is less obvious and can be quite misleading.
For instance, it can be zero if there is conditional dependence between two random variables and, even worse, its absolute value can be 
{arbitrarily close to}
one if two random variables are conditionally independent.
The partial copula is a natural generalization of the partial correlation coefficient and gives a meaningful measure of conditional dependence for general distributions. 
Moreover, there is a large class of distributions where the partial copula completely characterizes the conditional dependence. 

We first motivate and define the partial copula in \autoref{partial_copula} and then turn to some examples in \autoref{partial_examples}.
In \autoref{partial_cop_properties} we take a closer look at the properties of the partial copula. 
In particular, we examine the optimality of the partial copula as an approximation of the conditional copula and investigate its relation to conditional independence. \autoref{dependence_measures} considers dependence measures of the partial copula and how they are related to expected dependence measures of the conditional copula. 

%% file: partial_copula.tex
For simplicity, we consider continuous real-valued random variables with a joint positive density and assume that $\expec[Y_i]=0$ for $i=1,2$.
$Y_1\perp Y_2|Z$ means that $Y_1$ and $Y_2$ are independent given $Z$ and $C^{\perp}$ denotes the bivariate product copula.
For $i=1,2$, let $\beta_i: = \expec[ZZ']^{-1}\expec[Z'Y_i]$ so that
$Z\beta_i$ denotes the best linear predictor of $Y_i$ in terms of $Z,$  and define
\begin{align*}
{\cal E}_1 := Y_1 -Z\beta_1,
\quad &\text{and} \quad
\E_2 := Y_2 -Z\beta_2.
\end{align*}
The partial correlation of $Y_1$ and $Y_2$ given $Z$ can  be expressed as
\begin{align*}
\rho_{Y_1,Y_2\ps Z} = \corr[{\cal E}_1,{\cal E}_2].
\end{align*}
Thus, $\rho_{Y_1,Y_2\ps Z}$ is the correlation of $Y_1$ and $Y_2$ when each random variable has been corrected for the linear influence of $Z$, i.e.,
{$\corr[\tilde{g}(\E_i),\tilde{h}(Z)]=0$ for all linear functions $\tilde{g}$ and $\tilde{h}$.}
If $(Y_{1:2},Z)$ is jointly elliptically distributed, it is well known that $\expec[{\cal E}_1|Z]=\expec[{\cal E}_2|Z]=0$ (a.s.), implying that  $\rho_{Y_1,Y_2\ps Z}$ describes the dependence of $Y_1$ and $Y_2$ when their expectation does not depend on $Z$ anymore.
Moreover, if the elliptical distribution is a Gaussian distribution, $\rho_{Y_1,Y_2\ps Z}$ completely characterizes the conditional dependence.
One reason for that is that $\E_1\perp Z$ and $\E_2\perp Z$, which is equivalent to $\corr[h(\E_i),g(Z)]=0$ for $i=1,2$, and all measurable functions $h$ and $g$, so that $\rho_{Y_1,Y_2\ps Z}$ can be interpreted as the correlation of random variables which are individually independent of $Z$. 
In order to generalize this idea to the non-Gaussian case and to obtain random variables which are individually independent of $Z$, we define 
\begin{align*}
U_1 := F_{Y_1|Z}(Y_1|Z),
\quad &\text{and} \quad
U_2 := F_{Y_2|Z}(Y_2|Z).
\end{align*}
$U_i$ is called the conditional probability integral transform (CPIT) of $Y_i$ wrt $Z$. 
If $(Y_{1:2},Z)$ has a Gaussian distribution then $U_i = \Phi(\E_i/\sigma_{\E_i})$ so that $\corr[\Phi^{-1}(U_1),\Phi^{-1}(U_2)]=\rho_{Y_1,Y_2\ps Z}$.
However, even when the joint distribution of $(Y_{1:2},Z)$ is not Gaussian, we have that
$U_1\perp Z$ and $U_2\perp Z$ (Proposition 2.1 in \citet{Spanhel2015}). 
Thus, dependence measures which are based on the distribution of $(U_1,U_2)$
are also meaningful if the underlying distribution is not Gaussian. 
The joint distribution of the CPITs is the partial copula $\parsign{C}_{Y_1,Y_2\ps Z}$ of $F_{Y_1,Y_2|Z}$ and has been introduced by \citet{Bergsma2011}, \citet{Gijbels2015} and \citet{Spanhel2015}.%
\footnote{\citet{Bergsma2011} uses the partial copula to test for conditional independence. \citet{Gijbels2015} propose a non-parametric estimator of  the partial copula. \citet{Spanhel2015} show that partial copulas are optimal in the second tree of simplified vine copula approximations regarding the stepwise Kullback-Leibler divergence minimization.}

In some special cases, e.g., the Gaussian or Student-t distribution \citep{Stober2013b},
the partial copula and the conditional cdfs
$(\parsign{C}_{Y_1,Y_2\ps Z},F_{Y_1|Z},F_{Y_2|Z})$
determine $F_{Y_1,Y_2|Z}$ via 
$F_{Y_1,Y_2|Z}(y_1,y_2|z) = \parsign{C}_{Y_1,Y_2\ps Z}(u_1,u_2)$, where
$u_i = F_{Y_i|Z}(y_i|z), i = 1,2$.
However, in general, we have that
$F_{Y_1,Y_2|Z}(y_1,y_2|z) = C_{Y_1,Y_2|Z}(u_1,u_2|z): = \P(U_1\leq u_1,U_2\leq u_2|Z=z)$,
where $C_{Y_1,Y_2|Z}$ denotes the conditional copula of $F_{Y_1,Y_2|Z}$ \citep{Patton2006b}. 
Thus, the partial copula, which is  identical to the expected conditional copula $\int C_{Y_1,Y_2\ps Z}(\cdot,\cdot|t)dF_Z(t)$,  often acts as an approximation of $C_{Y_1,Y_2|Z}$ or $F_{Y_1,Y_2|Z}$.
At first sight, it might appear that the partial copula is a rather rough approximation of $C_{Y_1,Y_2|Z}$ since one assumes that $(U_1,U_2)$ are jointly independent of $Z$. 
However, by construction we have that $U_1\perp Z$ and $U_2 \perp Z$, so these necessary conditions for joint independence are satisfied. 
In particular,  $F_{Y_1,Y_2|Z}$ can be recovered from 
$(\parsign{C}_{Y_1,Y_2\ps Z},F_{Y_1|Z},F_{Y_2|Z})$
if and only if $Y_1$ and $Y_2$ can depend on $Z$ but the remaining dependence
between $U_1$ and $U_2$, which are individually independent of $Z$,
  does not depend on $Z$.
But also when $\parsign{C}_{Y_1,Y_2\ps Z}$ just acts as an approximation of $C_{Y_1,Y_2|Z}$, it is an attractive dependence measure because it is easier to estimate than $C_{Y_1,Y_2|Z}$ and measures conditional dependence by one bivariate unconditional copula and not by infinitely many bivariate unconditional copulas as it is the case for $C_{Y_1,Y_2|Z}$.
In the following, we give some explicit examples of  partial copulas.
Moreover, we investigate to what extent the approximation of $C_{Y_1,Y_2|Z}$ by $\parsign{C}_{Y_1,Y_2\ps Z}$ and the approximation of $F_{Y_1,Y_2|Z}$ by $(\parsign{C}_{Y_1,Y_2\ps Z},F_{Y_1|Z},F_{Y_2|Z})$  is optimal and examine properties of the partial copula and related dependence measures.

%% file: partial_examples.tex
\begin{myex}
[Trivariate FGM copula]
\label{fgm_ex}
The three-dimensional Farlie-Gumbel-Morgenstern (FGM) copula is given by
\begin{align*}
C_{1:3}(u_{1:3};\theta)
& = \prod_{i=1}^{3}u_i+\theta
\prod_{i=1}^{3}u_i(1-u_i),\quad |\theta|\leq 1.
\end{align*}
Since this copula is exchangeable, all three conditional copulas 
$C_{13\cs 2},C_{12\cs 3},C_{23\cs 1}$ are identical. 
It is straightforward to show that
\begin{align*}
C_{13\cs 2}(u_1,u_2|u_2) & = 
\prod_{i=1,3}u_{i|2}+\theta(1-2u_2)
\prod_{i=1,3}u_{i|2}(1-u_{i|2}),
\shortintertext{and}
\parsign{C}_{13\ps 2} &= C^\perp.
\end{align*}
\end{myex}

\begin{myex}[Trivariate Frank copula]
\label{frank_ex}
Consider the exchangeable three-dimensional Frank copula with dependence parameter $\theta> 0$,
\begin{align*}
C_{1:3}(u_{1:3}) & =
\log
\left\{
1-(1-\alpha)\prod_{i=1}^{3}\frac{1-\alpha^{u_i}}{1-\alpha}
\right\}\Big/\log(\alpha), 
\quad \alpha := \exp(-\theta)
.
\end{align*}
The conditional copula \citep{Mesfioui2008} belongs to the Ali-Mikhail-Haq (AMH) family with dependence parameter 
$\gamma(u_2;\theta) = 1-\exp(-\theta u_2)$, i.e.,
\begin{align*}
C_{13\cs 2}(u_1,u_2|u_2)
& = \frac{u_1u_2}{1-\gamma(u_2;\theta)
\prod_{i=1,3}(1-u_{i|2})}
.
\end{align*}
In \appref{Derive_Frank_ex} the closed-form expression for the partial copula is derived, which is given by
\begin{align*}
\parsign{C}_{13\ps 2}(u_1,u_2) & = 
\frac{u_1u_2}{\theta f(u_1,u_2)}
\left[
\log\big(1-(1-\exp(-\theta))(1-f(u_1,u_2))\big)+\theta
\right]\!,
\end{align*}
where $f(u_1,u_2)
:=u_1+u_2-u_1u_2$.
\autoref{FigPartialFrank} illustrates $C_{13|2}$ and $\parsign{C}_{13\ps 2}$.
\end{myex}

\begin{myex}
[Partial copulas of the Gaussian, Student-t, and Clayton copula]
\label{ex_SimplifiedDistr}
For the Gaussian, Student-t, and Clayton copula, conditional  and partial copulas coincide since the simplifying assumption holds \citep{Stober2013b}.
\end{myex}

%% file: fig_partial_frank.tex
\begin{figure}[h!]
\centering
\includegraphics[scale=0.8]{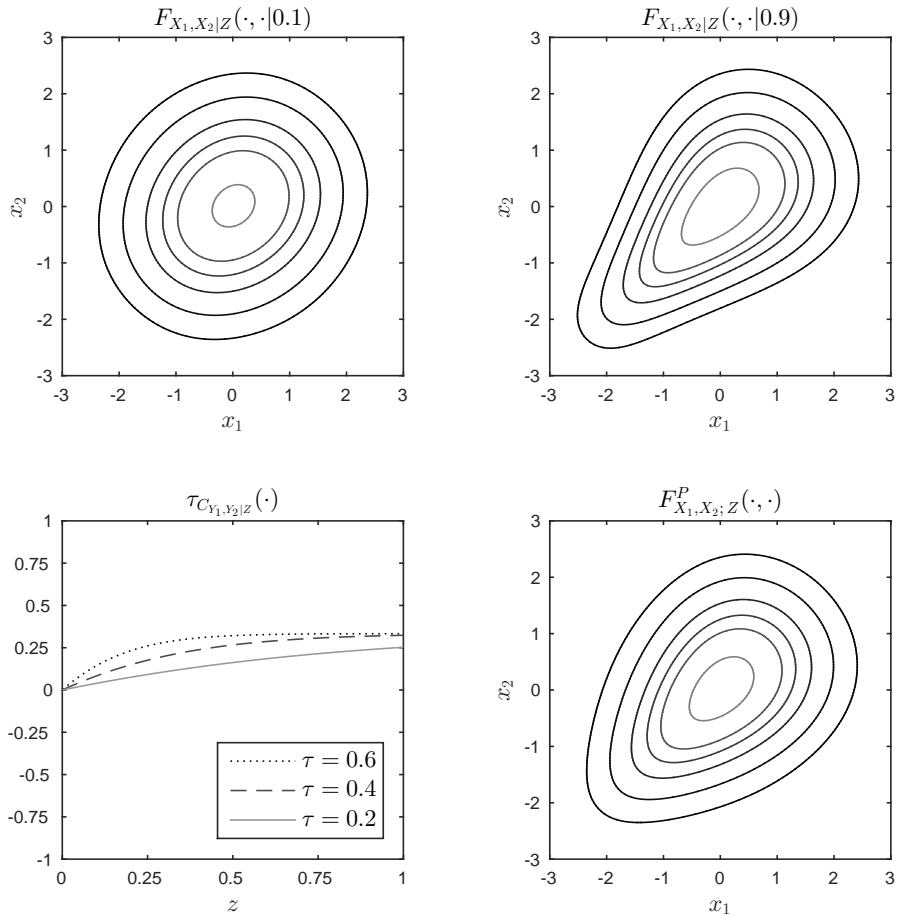}
{\scriptsize
\caption{Illustration of the conditional and partial copula if $(Y_{1:2},Z)$ is distributed according to a trivariate Frank copula with pairwise values of Kendall's $\tau$ being $0.4$. The upper panel shows contour plots of the density of $F_{X_{1},X_{2}|Z}(\cdot,\cdot|z)$ where
 $X_{i}:=\Phi^{-1}(U_{i}) = \Phi^{-1}(F_{Y_{i}|Z}(Y_{i}|Z))$, $i=1,2$, and $\Phi$ is the cdf of the standard normal distribution.
On the left hand side in the lower panel the variation of Kendall's $\tau$ of $C_{Y_{1},Y_{2}\cs Z}$ is depicted. The lower right figure shows contour plots of the density of $\parsign{F}_{X_{1},X_{2}\ps Z}$.}
\label{FigPartialFrank}
}
\end{figure}

%% file: partial_cop_properties.tex
\newcommand{\ltwodist}[2]{D_{L2}(#1,#2)}
\newcommand{\paraitfgm}{(\prod_{k=2}^{K-1}u_{k})}%
\renewcommand{\paraitfgm}{u_2}
As stated in \autoref{partial_copula}, the partial copula can be considered as an approximation of the conditional copula. The first property shows that the partial copula minimizes the Kullback-Leibler (KL) divergence from the conditional copula.
\begin{myprop}
[KL divergence minimization]
\label{kl_min}
The partial copula $\parsign{C}_{Y_1,Y_2\ps Z}$ minimizes the KL divergence from the conditional copula $C_{Y_1,Y_2|Z}$  in the space of absolutely continuous bivariate distribution functions.
\end{myprop}
\begin{myproof}
This follows from equation (3.3) in Theorem 3.1 in \citet{Spanhel2015}.
\end{myproof}
However, the partial copula does not always minimize the $L^2$ distance to the conditional copula.

\begin{myprop}
[$L^2$ distance minimization]
\label{l2distance}
Let $C_{Y_1,Y_2|Z}(U_1,U_2|Z)$ have finite variance and ${\cal F}_2$ be the space of bivariate cdfs so that each $C_{Y_1,Y_2\ps Z}(U_1,U_2)\in{\cal F}_2$ is a $C_{Y_1,Y_2|Z}$-measurable random variable with finite variance.
Let $C^{L^2}_{Y_1,Y_2\ps Z}\in{\cal F}_2$ denote the bivariate cdf which minimizes the $L^2$ distance to $C_{Y_1,Y_2|Z}$. 
In general,
\begin{align*}
C^{L^2}_{Y_1,Y_2\ps Z}  & \neq \parsign{C}_{Y_1,Y_2\ps Z}.
\end{align*}
\end{myprop}
\begin{myproof}
See \appref{proof_l2distance}.
\end{myproof}
The first two properties address the optimality of the partial copula when it comes to approximating conditional copulas. 
But partial copulas, together with univariate conditional cdfs, can also be used to provide a model for a general bivariate conditional distribution. 
For instance, \citet{Song2009} use generalized linear models for univariate conditional cdfs and join these conditional cdfs with an unconditional copula.
Also, the conditional cdfs of financial returns are often filtered with ARMA-GARCH models and the remaining dependence is then modeled by an unconditional copula \citep{Chen2006,Liu2009,Min2014,Nikoloulopoulos2012}.
Therefore, the next property is of interest if conditional cdfs are linked with the partial copula. 

\begin{myprop}[Non-optimality of partial copula-induced approximations]
\label{cop_approx}
Let 
\begin{align*}
\parsign{F}_{Y_1,Y_2|Z}(y_1,y_2|z):= 
\parsign{C}_{Y_1,Y_2\ps Z}\big(F_{Y_1|Z}(y_1|z),F_{Y_2|Z}(y_2|z)\big),
\end{align*}
be the approximation of  $F_{Y_1,Y_2|Z}$ that emerges if the conditional copula is approximated by the partial copula and the true univariate conditional cdfs. 
In general, $\parsign{F}_{Y_1,Y_2|Z}$ does neither minimize the KL divergence nor the $L^2$ distance from 
${F}_{Y_1,Y_2|Z}$.
\end{myprop}
\begin{myproof}
See \appref{proof_cop_approx}.
\end{myproof}
Although the partial copula $\parsign{C}_{Y_1,Y_2\ps Z}$ minimizes the KL divergence from ${C}_{Y_1,Y_2\cs Z}$ (\autoref{kl_min}),
\autoref{cop_approx} reveals the surprising result that this does not imply that the induced approximation $\parsign{F}_{Y_1,Y_2|Z}$ also minimizes the KL divergence from ${F}_{Y_1,Y_2|Z}$.
Note that \autoref{kl_min} implies that $\parsign{C}_{Y_1,Y_2\ps Z}$ is the bivariate copula that minimizes the KL divergence if the true conditional cdfs $(F_{Y_1|Z},F_{Y_2|Z})$ are specified. 
Thus, \autoref{cop_approx} implies that, in general, one can obtain a better approximation by an adequate misspecification of the conditional cdfs.
That is, there are conditional cdfs $(\tilde{F}_{Y_1|Z},\tilde{F}_{Y_2|Z})$ such that $(\tilde{F}_{Y_1|Z},\tilde{F}_{Y_2|Z})\neq ({F}_{Y_1|Z},{F}_{Y_2|Z})$
and $\expec[\parsign{C}_{Y_1,Y_2\ps Z}\big(\tilde{F}_{Y_1|Z}(Y_1|Z),
\tilde{F}_{Y_2|Z}(Y_2|Z)\big)]>\expec[\parsign{F}_{Y_1,Y_2|Z}(Y_1,Y_2|Z)] $.
Because the marginal distributions of $(\tilde{F}_{Y_1|Z}(Y_1|Z),
\tilde{F}_{Y_2|Z}(Y_2|Z))$ are not uniform in general,
one can further improve the approximation if one specifies a pseudo-copula \citep{Fermanian2012}. 
Another interesting implication of \autoref{cop_approx} is that, if the conditional cdfs and the partial copula are estimated, 
the joint and stepwise ML estimator may have a different probability limit  if the partial and conditional copula do not coincide.

\begin{myprop}[Joint and stepwise ML estimation]
\label{joint_seq}
Wlog assume that the following random variables and parameters are scalars.
Let $(F_{Y_1|Z},F_{Y_2|Z})$ be the true conditional cdfs and $\parsign{C}_{Y_1,Y_2\ps Z}$ be the true partial copula of $C_{Y_1,Y_2|Z}$. 
Assume that we observe $n$ independent samples from $(Y_{1:2},Z)$.
For $i=1,2,$ let $\tilde{F}_{Y_i|Z}(\tilde{\theta}_i)$ be a parametric conditional cdf and assume that
$\exists \theta_i\in\Theta_i\colon \tilde{F}_{Y_i|Z}(\theta_i)=F_{Y_i|Z}$. 
Let $\tilde{C}_{12}(\tilde{\theta}_3)$ be a parametric bivariate copula family  so that $\exists \theta_3\in\Theta_3\colon\tilde{C}_{12}(\theta_3)=\parsign{C}_{Y_1,Y_2\ps Z}$.
Let
\begin{align*}
\theta^{J}_n := \argmax_{\tilde\theta_{1:3}\in\Theta_{1:3}}
\sum_{i=1}^n \log
\Big(\tilde{c}_{12}
\big(\tilde{F}_{Y_1|Z}(Y_{1,i}|Z_{i};\tilde\theta_1),
\tilde{F}_{Y_2|Z}(Y_{2,i}|Z_{i};\tilde\theta_2);\tilde\theta_3\big)
\prod_{j=1}^2\tilde{f}_{Y_j|Z}(Y_{j,i}|Z_{i};\tilde\theta_j)\Big)
\end{align*}
be the joint ML estimator and 
\begin{align*}
\theta^{S}_n := 
\begin{pmatrix}
\theta_{1:2}^S\\ \theta_3^{S}
\end{pmatrix}
= 
\begin{pmatrix}
\displaystyle
\argmax_{\tilde\theta_{1:2}\in\Theta_{1:2}}
\sum_{j=1}^2\sum_{i=1}^n \log 
\tilde{f}_{Y_j|Z}(Y_{j,i}|Z_{i};\tilde\theta_j)
\\
\displaystyle
\argmax_{\tilde\theta_3\in\Theta_3}\sum_{i=1}^n \log
\tilde{c}_{12}
\big(\tilde{F}_{Y_1|Z}(Y_{1,i}|Z_{i};\theta_1^{S}),\tilde{F}_{Y_2|Z}(Y_{2,i}|Z_{i};\theta_2^{S})
;\tilde\theta_3\big)
\end{pmatrix}
\end{align*}
be the stepwise ML estimator. 
Assume that the regularity conditions stated in \citep{Joe2005} hold and that
\begin{align*}
\gamma := \argmax_{\tilde\theta_{1:3}\in\Theta_{1:3}} 
\expec\Big[
\log 
\big(\tilde{c}_{12}
\big(\tilde{F}_{Y_1|Z}(Y_{1}|Z;\tilde\theta_1),
\tilde{F}_{Y_2|Z}(Y_{2}|Z;\tilde\theta_2);\tilde\theta_3\big)
\prod_{j=1}^2\tilde{f}_{Y_j|Z}(Y_{j}|Z;\tilde\theta_j)\big)\Big]
\end{align*}
exists.
If $C_{Y_1,Y_2|Z}=\parsign{C}_{Y_1,Y_2\ps Z}$ (a.s.), then 
$\theta_{n}^J
\stackrel{p}{\to} \theta$
and 
$\theta_{n}^S
\stackrel{p}{\to} \theta$ for $n\to\infty$. 
However, if $C_{Y_1,Y_2|Z}\neq \parsign{C}_{Y_1,Y_2\ps Z}$ (a.s.), then 
$\theta_n^J\stackrel{p}{\to} \gamma$ and $\theta_n^S\stackrel{p}{\to} \theta$ for $n\to\infty$.
In particular, $\theta_{i,n}^J-\theta_{i,n}^S
\stackrel{p}{\to} 0$ may \underline{not} hold for all $i=1,2,3$.
\end{myprop}
\begin{myproof}
See \appref{proof_joint_seq}.
\end{myproof}
Thus, the well known result that the joint and stepwise ML estimator of conditional cdfs and an unconditional copula converge to the same probability limit may not hold if the partial and conditional copula do not coincide. 
The next two properties are related to the parametric family of the partial copula.
\begin{myprop}[Archimedean copulas]
\label{not_archimedean}
Let the copula of $(Y_{1:2},Z)$ be Archimedean. 
Then $\parsign{C}_{Y_1,Y_2\ps Z}$ might not be Archimedean.
\end{myprop}
\begin{myproof}
See \appref{proof_not_archimedean}.
\end{myproof}
Note that, if the  copula of $(Y_{1:2},Z)$ is Archimedean, $C_{Y_1,Y_2|Z}$ is always Archimedean \citep{Mesfioui2008}.

\begin{myprop}[Family of the partial copula]
Assume that there is a bivariate parametric copula family $C^F(\cdot,\cdot;\theta)$ with parameter $\theta\in\Theta$ and a measurable function $g$ such that 
$C_{Y_1,Y_2|Z}(\cdot,\cdot,|z) = C^F(\cdot,\cdot;g(z))$ for almost all $z$.
In general, it does not hold that $\exists\theta\in\Theta: \parsign{C}_{Y_1,Y_2\ps Z}(\cdot,\cdot)=
C^F(\cdot,\cdot,\theta).$
\end{myprop}
\begin{myproof}
This follows from \autoref{not_archimedean}.
\end{myproof}
Thus, partial copulas can also be used to obtain new (unconditional) copulas (see \autoref{frank_ex}).
One deficiency of the partial correlation coefficient is that its absolute value can be arbitrarily close to one if we have conditional independence.
The partial copula is more attractive in this regard as the following property demonstrates.

\begin{myprop}[Conditional (in)dependence]
\label{cond_indep}
Let $Y_1\perp Y_2|Z$. {The smallest upper bound for the absolute value of $\rho_{Y_1,Y_2\ps Z}$ is one.}
However, we always have that $\parsign{C}_{Y_1,Y_2\ps Z}=C^{\perp}$.
On the other side, $\parsign{C}_{Y_1,Y_2\ps Z}=C^{\perp}$ or $\rho_{Y_1,Y_2\ps Z}=0$ does not imply that $Y_1\perp Y_2|Z$.
\end{myprop}
\begin{myproof}
See \appref{proof_cond_indep}.
\end{myproof}
The next property points out that a varying conditional correlation is not sufficient for the non-equality of the partial and conditional copula.

\begin{myprop}[Conditional correlation]
\label{cond_corr}
If $\rho_{Y_1,Y_2|Z}(Z)=(\expec[Y_1Y_2|Z]-\expec[Y_1|Z]\expec[Y_2|Z])/
\sqrt{Var[Y_1|Z]Var[Y_2|Z]}$ is not almost surely a constant this does not imply that 
$\P( C_{Y_1,Y_2|Z}= \parsign{C}_{Y_1,Y_2\ps Z})<1$.
\end{myprop}
\begin{myproof}
See \appref{proof_cond_corr}.
\end{myproof}

%% file: dependence_measures.tex
Let $\rho_{\parsign{C}_{Y_1,Y_2\ps Z}}, \tau_{\parsign{C}_{Y_1,Y_2\ps Z}},
\lambda_{\parsign{C}_{Y_1,Y_2\ps Z}}^l,$ and
$\lambda_{\parsign{C}_{Y_1,Y_2\ps Z}}^{u}$, denote Spearman's $\rho$, Kendall's $\tau$, and the corresponding lower and upper tail dependence coefficients of $\parsign{C}_{Y_1,Y_2\ps Z}$. We refer to these dependence measures as partial dependence measures.
The next proposition summarizes that partial Spearman's $\rho$ and the partial  tail dependence coefficients are equal to the expected Spearman's $\rho$ and the expected tail dependence coefficients of the conditional copula.
\begin{myprop}[Partial Spearman's $\rho$ and partial tail dependence]
\label{exp_cond}
It holds that
\begin{align*}
\rho_{\parsign{C}_{Y_1,Y_2\ps Z}} & = \expec[\rho_{{C}_{Y_1,Y_2| Z}}(Z)]
\\
\lambda_{\parsign{C}_{Y_1,Y_2\ps Z}}^l & = 
\expec[\lambda_{{C}_{Y_1,Y_2| Z}}^l(Z)]
\\
\lambda_{\parsign{C}_{Y_1,Y_2\ps Z}}^{u} & = 
\expec[\lambda_{{C}_{Y_1,Y_2| Z}}^{u}(Z)]
\end{align*}
\end{myprop}
\begin{myproof}
These statements are easily verified by computing the expectations.
\end{myproof}
However, the expected conditional Kendall's $\tau$ is in general not equal to partial Kendall's $\tau$.%
\footnote{This has also been remarked by Ir{\`e}ne Gijbels during her {t}alk
``Nonparametric testing for no covariates effects in conditional copulas'' at the nonparametric copula day in {M}unich, 2015.}
\begin{myprop}
[Partial Kendall's $\tau$]
\label{kendall}
In general,
$
\tau_{\parsign{C}_{Y_1,Y_2\ps Z}}\neq \expec[\tau_{C_{Y_1,Y_2|Z}}\left(Z\right)].
$
\end{myprop}
\begin{myproof}
See \appref{proof_kendall}.
\end{myproof}
Unless 
$|\rho_{\parsign{C}_{Y_1,Y_2\ps Z}}|=1$, the value of $\rho_{\parsign{C}_{Y_1,Y_2\ps Z}}$ does not provide any information about the value of $\rho_{{C}_{Y_1,Y_2| Z}}$.
E.g., $\rho_{\parsign{C}_{Y_1,Y_2\ps Z}}=0$ does not imply that 
$\rho_{{C}_{Y_1,Y_2| Z}}=0$ (a.s.)
(see \autoref{fgm_ex}).
However, for the coefficients of tail dependence we obtain the following relation.
\begin{myprop}
[Tail dependence]
$\lambda_{\parsign{C}_{Y_1,Y_2\ps Z}}^{l}=0 \eq 
\lambda_{{C}_{Y_1,Y_2| Z}}^l(Z)=0$ (a.s)
and 
$\lambda_{\parsign{C}_{Y_1,Y_2\ps Z}}^{u}=0 \eq \lambda_{{C}_{Y_1,Y_2| Z}}^{u}(Z)=0$ (a.s.). 
\end{myprop}
\begin{myproof}
Follows by \autoref{exp_cond} and because the coefficients of tail dependence are non-negative.
\end{myproof}
Thus, the partial copula has no tail dependence 
if and only if the conditional copula has no tail dependence (\as).
This is a useful result because we can test for tail dependence of the conditional copula by testing tail dependence of the partial copula. For instance, when modeling a time-varying conditional copula it is important to know whether the time-varying conditional copula should exhibit tail dependence.

%% file: conclusion.tex
The partial copula is a natural generalization of the partial correlation coefficient which does not share some of its drawbacks.
We presented  examples of the partial copula and investigated several of its properties. 
The bivariate partial copula minimizes the KL divergence to a bivariate conditional copula but the resulting approximation of a general bivariate conditional cdf does in general not minimize the KL divergence. 
As a result, the joint and stepwise ML estimator may converge to different probability limits.

While the partial copula has attractive theoretical properties, its estimation is much more involved than the estimation of the partial correlation coefficient. Non-parametric estimation of the partial copula has been proposed by \citet{Gijbels2015} if there is only one conditioning variable. 
However, further investigation is required to determine whether such a non-parametric estimator is reasonable if the set of conditioning variables is not very small and we have finite sample sizes.
Alternatively, one could use higher-order partial copulas \citep{Spanhel2015} which constitute a different generalization of the partial correlation coefficient. While higher-order partial copulas do not share all properties of partial copulas, they can be efficiently estimated for a very large set of conditioning variables.

%% file: proof_partial_frank.tex
\subsection{Derivation of the partial Frank copula (\autoref{frank_ex})}\label{Derive_Frank_ex}
Let $g(u_1,u_2,\theta) 
:= 1-\left(1-\exp\left(-\theta\right)\right)\left(1-u_1\right)\left(1-u_2\right)$
and $f(u_1,u_2) := u_1+u_2-u_1u_2$.
The partial copula is given by
\begin{align*}
\parsign{C}_{13\ps 2}\left(u_1,u_2|u_2\right) &= \int_{0}^{1} C_{13\cs 2}\left(u_1,u_2|u_2\right) \text{d}u_2 
= \int_{0}^{1} \frac{u_1u_2}{1-\left
(1-\exp\left(-\theta u_2\right)\right)\left(1-u_1\right)\left(1-u_2\right)} \text{d}u_2 \\
&= \int_{1}^{g(u_1,u_2,\theta)} \frac{u_1u_2}{x\theta\left(f(u_1,u_2)-x\right)} \text{d}x 
= \frac{u_1u_2}{\theta f(u_1,u_2)} \int_{1}^{g(u_1,u_2,\theta)} \frac{1}{x} - \frac{1}{x-f(u_1,u_2)} \text{d}x \\
&= \frac{u_1u_2}{\theta f(u_1,u_2)} \left[\log\left(x\right) - \log
\left(x- f(u_1,u_2)\right) \right]_{x=1}^{x=g(u_1,u_2,\theta)} 
= \frac{u_1u_2}{\theta f(u_1,u_2)} 
\left[\log g(u_1,u_2,\theta)+\theta\right].
\end{align*}

%% file: proof_l2distance.tex
\subsection{Proof of \autoref{l2distance}}
\label{proof_l2distance}
It is well known  that, if the variance of $Y$ exists,  $\expec[Y|X]$ minimizes the $L^2$ distance to $Y$ over all $X$-measurable random variables with finite variance. 
Thus
\begin{align*}
C^{L2}_{Y_1,Y_2\ps Z}(U_1,U_2) & =   \std{\arg\min}{C_{Y_1,Y_2\ps Z}\in {\cal F}_2}
\expec\big[(C_{Y_1,Y_2|Z}(U_1,U_2|Z)-C_{Y_1,Y_2\ps Z}(U_1,U_2))^2\big]
= \expec\big[C_{Y_1,Y_2|Z}(U_1,U_2|Z)|U_1,U_2\big].
\end{align*}
It is easy to check that $C^{L^2}_{Y_1,Y_2\ps Z}$ is a bivariate copula. If $C_{Y_1,Y_2,Z}$  is the FGM copula given in \autoref{fgm_ex} then
\begin{align*}
C_{Y_1,Y_2\ps Z}^{L^2}(u_1,u_2) & = 
\int_0^1C_{Y_1,Y_2|Z}(u_1,u_2|z)f_{Z|U_1,U_2}(z|u_1,u_2)dz
\\
& = 
u_1u_2\left(1+\frac{\theta^2(4u_1^2u_2^2-6(u_1^2u_2+u_1u_2^2)
+2(u_1^2+u_2^2)+9u_1u_2-3(u_1+u_2)+1)}{3}\right) 
\\
& \neq u_1u_2 = \parsign{C}_{Y_1,Y_2\ps Z}(u_1,u_2).
\end{align*}

%% file: proof_cop_approx.tex
\subsection{Proof of \autoref{cop_approx}}
\label{proof_cop_approx}
Wlog assume that $Z$ is a scalar and that $(Y_1,Y_2,Z)$ is a uniform random vector with $(Y_1,Z)\sim C_{12}$ and $(Z,Y_2)\sim C_{23}$. 
The KL divergence of $\tilde{F}_{Y_1,Y_2|Z}$ from $F_{Y_1,Y_2|Z}$ is given by 
\begin{align*}
D_{KL}(F_{Y_1,Y_2|Z},\tilde{F}_{Y_1,Y_2|Z}) & =
\expec\left[\log\frac{f_{Y_1,Y_2|Z}(Y_1,Y_2|Z)}{\tilde{f}_{Y_1,Y_2|Z}(Y_1,Y_2|Z)}\right]
\end{align*}
and is identical to the KL divergence given in equation (3.1) in \citet{Spanhel2015}.
From equation (3.7) in Theorem 3.2 in \citet{Spanhel2015} it  follows that $D_{KL}(F_{Y_1,Y_2|Z},\tilde{F}_{Y_1,Y_2|Z})$ does, in general, not attain a minimum at $\tilde{F}_{Y_1,Y_2|Z}= \parsign{F}_{Y_1,Y_2|Z}$.
That $\parsign{F}_{Y_1,Y_2|Z}$ may not minimize the $L^2$ distance to $F_{Y_1,Y_2|Z}$ follows from \autoref{l2distance}.

%% file: proof_joint_seq.tex
\subsection{Proof of \autoref{joint_seq}}
\label{proof_joint_seq}
\citet{Joe2005} establishes regularity conditions so that, if $C_{Y_1,Y_2|Z}=\parsign{C}_{Y_1,Y_2\ps Z}$ (a.s.), then $\theta_n^{S}\stackrel{p}{\to}\theta$ and $\theta_n^{J}\stackrel{p}{\to}\theta$, which implies that $\theta_n^J-\theta_n^S \stackrel{p}{\to} 0$ holds.
Now, let $C_{Y_1,Y_2|Z}\neq \parsign{C}_{Y_1,Y_2\ps Z}$ and assume that 
$\gamma = \argmax_{\tilde\theta_{1:3}\in\Theta_{1:3}} 
\expec\big[
\log 
\big(\tilde{c}_{12}
\big(\tilde{F}_{Y_1|Z}(Y_{1}|Z;\tilde\theta_1),
\tilde{F}_{Y_2|Z}(Y_{2}|Z;\tilde\theta_2);\tilde\theta_3\big)
\prod_{j=1}^2\tilde{f}_{Y_j|Z}(Y_{j}|Z;\tilde\theta_j)\big)\big]
$
exists.
Note that,
$\displaystyle\theta_{1:2} = \argmax_{\tilde{\theta}_{12}\in\Theta_{12}}
\expec\big[\sum_{j=1}^2\log\tilde{f}_{Y_j|Z}(Y_{j}|Z;\tilde\theta_j)\big]$
and 
$\displaystyle\theta_3 = \argmax_{\tilde\theta_3\in\Theta_3} 
\expec\big[\log\tilde{c}_{12}\big(\tilde{F}_{Y_1|Z}(Y_{1}|Z;\theta_1),
\tilde{F}_{Y_2|Z}(Y_{2}|Z;\theta_2)
;\tilde\theta_3\big)\big]$.
Thus, under the regularity conditions in \citet{Joe2005}, it follows that $\theta^S_n\stackrel{p}{\to}\theta$.
\autoref{cop_approx} and the subsequent remarks imply that, in general, $\gamma_i\neq \theta_i$ for all $i=1,2,3$. 
Provided the regularity conditions in \citet{Joe2005} are satisfied, it follows that $\theta^{J}_{i,n}\stackrel{p}{\to} \gamma_i$ for all $i=1,2,3$, which finishes the proof.

%% file: proof_not_archimedean.tex
\subsection{Proof of \autoref{not_archimedean}}
\label{proof_not_archimedean}
Let $\parsign{C}_{Y_1,Y_2\ps Z}$ be given as in \autoref{frank_ex}.
We observe that
\begin{align*}
\parsign{C}_{13\ps 2}\big(\parsign{C}_{13\ps 2}(0.25,0.5),0.5\big) \neq 
\parsign{C}_{13\ps 2}\big(0.25,\parsign{C}_{13\ps 2}(0.5,0.5)\big),
\end{align*}
which shows that $\parsign{C}_{13\ps 2}$ is not associative, which is sufficient for the copula not to be Archimedean \citep{Nelsen2006}.

%% file: proof_cond_indep.tex
\subsection{Proof of \autoref{cond_indep}}
\label{proof_cond_indep}
Let $Z\sim N(0,1)$ and  $Y_1 = -1+ Z^2+\E_1, Y_2 = -1+Z^2+\E_2$, where $\E_1, \E_2$, and $Z$ are mutually independent, $\expec[\E_i]=0$ and 
$\var[\E_i]=\sigma$ for $i=1,2$.%
\footnote{If $\sigma=0$, then $(Y_{1:2},Z)$
does not have a Lebesgue density, which we assume throughout the paper.
However, if we allow that $\E_i$ is almost surely a constant, it follows that the maximal absolute value of $\rho_{Y_1,Y_2\ps Z}$ is one.}
It is easy to show that $Y_1\perp Y_2|Z$.
Note that $\cov[Y_i,Z] = \expec[Z^3]=0$ so that
$Y_i-(1,Z)\beta_i = Z^2+\E_i$, where $\beta_i$ is the best linear predictor of $Y_i$ in terms of $Z$.
Thus, $\rho_{Y_1,Y_2\ps Z}=\corr[Z^2+\E_1,Z^2+\E_2] = \frac{\var[Z^2]}{\var[Z^2]+\sigma}$, and
 $\lim_{\sigma\to 0}\rho_{Y_1,Y_2\ps Z}=1$.
Setting $Y_2 = -Z^2+\E_2$ shows that $\lim_{\sigma\to 0}\rho_{Y_1,Y_2\ps Z}=-1$.
If $Y_1\perp Y_2|Z$ then 
$C_{Y_1,Y_2|Z}(U_1,U_2|Z) = U_1U_2$, i.e., $\parsign{C}_{Y_1,Y_2\ps Z}=C^{\perp}$.
From \autoref{fgm_ex} it follows that
 $\parsign{C}_{Y_1,Y_2\ps Z}=C^{\perp}$ does not imply that $Y_1\perp Y_2|Z$.

%% file: proof_cond_corr.tex
\subsection{Proof of \autoref{cond_corr}}
\label{proof_cond_corr}
Let $Z$ be exponentially distributed with unit mean,  
$Y_1|Z\sim \log {\cal N}(0,1)$ and $Y_2|Z=z\sim \log{\cal N}(0,z)$, where $\log{\cal N}(0,\sigma)$ denotes the log-normal distribution with zero location parameter and scale parameter $\sigma$. 
Using the same arguments as in Example 5.26 in \cite{McNeil2005}, we obtain that, if 
$C_{Y_1,Y_2|Z}(U_1,U_2|Z)=\min(U_1,U_2) = \parsign{C}_{Y_1,Y_2\ps Z}$,  the support of the random variable
\begin{align*}
\rho_{Y_1,Y_2|Z}(Z) = 
\frac{\exp(Z)-1}{\sqrt{(\exp(1)-1)(\exp(Z^2)-1)}}
\end{align*}
is the unit interval $[0,1]$.

%% file: proof_kendall.tex
\subsection{Proof of \autoref{kendall}}
\label{proof_kendall}
Let $Z$ be uniformly distributed and consider the following conditional copula 
\begin{align*}
C_{Y_1,Y_2|Z}\left(u_1,u_2|z\right) &= u_1u_2 + 
z u_1 u_2 \left(1-u_1\right) \left(1-u_2\right)(1+ u_1u_2).
\end{align*}
The corresponding partial copula is 
\begin{align*}
\parsign{C}_{Y_1,Y_2\ps Z}(u_1,u_2) &= u_1u_2 + \frac{1}{2}  u_1u_2 \left(1-u_1\right) \left(1-u_2\right)(1+ u_1u_2).
\end{align*}
Elementary integration yields that Kendall's $\tau$ for the conditional copula is given by
\begin{align*}
\tau_{C_{Y_1,Y_2|Z}}\left(Z\right) & = 
4 \int_{\left[0,1\right]^{2}} C_{Y_1,Y_2|Z}\left(u_1,u_2|Z\right) 
dC_{Y_1,Y_2|Z}\left(u_1,u_2|Z\right) - 1
 = Z^2/450 + 5Z/18 
\end{align*}
Thus, 
\begin{align*}
\expec[\tau_{C_{Y_1,Y_2|Z}}\left(Z\right)] &= \frac{377}{2700}
\neq \frac{251}{1800} =
\tau_{\parsign{C}_{Y_1,Y_2\ps Z}}.
\end{align*}

%% file: manuscript.bbl
\begin{thebibliography}{15}
\expandafter\ifx\csname natexlab\endcsname\relax\def\natexlab#1{#1}\fi
\providecommand{\url}[1]{\texttt{#1}}
\providecommand{\href}[2]{#2}
\providecommand{\path}[1]{#1}
\providecommand{\DOIprefix}{doi:}
\providecommand{\ArXivprefix}{arXiv:}
\providecommand{\URLprefix}{URL: }
\providecommand{\Pubmedprefix}{pmid:}
\providecommand{\doi}[1]{\href{http://dx.doi.org/#1}{\path{#1}}}
\providecommand{\Pubmed}[1]{\href{pmid:#1}{\path{#1}}}
\providecommand{\bibinfo}[2]{#2}
\ifx\xfnm\relax \def\xfnm[#1]{\unskip,\space#1}\fi
\bibitem[{{Bergsma}(2011)}]{Bergsma2011}
\bibinfo{author}{{Bergsma}, W.}, \bibinfo{year}{2011}.
\newblock \bibinfo{title}{{Nonparametric testing of conditional independence by
  means of the partial copula}}.
\newblock \bibinfo{journal}{ArXiv e-prints}
  \href{http://arxiv.org/abs/1101.4607}{\tt arXiv:1101.4607}.
\bibitem[{Chen and Fan(2006)}]{Chen2006}
\bibinfo{author}{Chen, X.}, \bibinfo{author}{Fan, Y.}, \bibinfo{year}{2006}.
\newblock \bibinfo{title}{{E}stimation and model selection of semiparametric
  copula-based multivariate dynamic models under copula misspecification}.
\newblock \bibinfo{journal}{Journal of Econometrics} \bibinfo{volume}{135},
  \bibinfo{pages}{125--154}.
\bibitem[{Fermanian and Wegkamp(2012)}]{Fermanian2012}
\bibinfo{author}{Fermanian, J.D.}, \bibinfo{author}{Wegkamp, M.H.},
  \bibinfo{year}{2012}.
\newblock \bibinfo{title}{{T}ime-dependent copulas}.
\newblock \bibinfo{journal}{Journal of Multivariate Analysis}
  \bibinfo{volume}{110}, \bibinfo{pages}{19--29}.
\bibitem[{Gijbels et~al.(2015)Gijbels, Omelka and Veraverbeke}]{Gijbels2015}
\bibinfo{author}{Gijbels, I.}, \bibinfo{author}{Omelka, M.},
  \bibinfo{author}{Veraverbeke, N.}, \bibinfo{year}{2015}.
\newblock \bibinfo{title}{Estimation of a copula when a covariate affects only
  marginal distributions}.
\newblock \bibinfo{journal}{Scandinavian Journal of Statistics}
  \bibinfo{volume}{42}, \bibinfo{pages}{1109--1126}.
\bibitem[{Joe(2005)}]{Joe2005}
\bibinfo{author}{Joe, H.}, \bibinfo{year}{2005}.
\newblock \bibinfo{title}{Asymptotic efficiency of the two-stage estimation
  method for copula-based models}.
\newblock \bibinfo{journal}{Journal of Multivariate Analysis}
  \bibinfo{volume}{94}, \bibinfo{pages}{401 -- 419}.
\bibitem[{Liu and Luger(2009)}]{Liu2009}
\bibinfo{author}{Liu, Y.}, \bibinfo{author}{Luger, R.}, \bibinfo{year}{2009}.
\newblock \bibinfo{title}{{E}fficient estimation of copula-{G}{A}{R}{C}{H}
  models}.
\newblock \bibinfo{journal}{Computational Statistics {\&} Data Analysis}
  \bibinfo{volume}{53}, \bibinfo{pages}{2284--2297}.
\bibitem[{McNeil et~al.(2005)McNeil, Frey and Embrechts}]{McNeil2005}
\bibinfo{author}{McNeil, A.J.}, \bibinfo{author}{Frey, R.},
  \bibinfo{author}{Embrechts, P.}, \bibinfo{year}{2005}.
\newblock \bibinfo{title}{{Q}uantitative risk management}.
\newblock {P}rinceton series in finance, \bibinfo{publisher}{Princeton Univ.
  Press}, \bibinfo{address}{Princeton NJ}.
\bibitem[{Mesfioui and Quessy(2008)}]{Mesfioui2008}
\bibinfo{author}{Mesfioui, M.}, \bibinfo{author}{Quessy, J.F.},
  \bibinfo{year}{2008}.
\newblock \bibinfo{title}{{D}ependence structure of conditional {A}rchimedean
  copulas}.
\newblock \bibinfo{journal}{Journal of Multivariate Analysis}
  \bibinfo{volume}{99}, \bibinfo{pages}{372--385}.
\bibitem[{Min and Czado(2014)}]{Min2014}
\bibinfo{author}{Min, A.}, \bibinfo{author}{Czado, C.}, \bibinfo{year}{2014}.
\newblock \bibinfo{title}{{S}{C}{O}{M}{D}{Y} models based on pair-copula
  constructions with application to exchange rates}.
\newblock \bibinfo{journal}{Computational Statistics {\&} Data Analysis}
  \bibinfo{volume}{76}, \bibinfo{pages}{523--535}.
\bibitem[{Nelsen(2006)}]{Nelsen2006}
\bibinfo{author}{Nelsen, R.B.}, \bibinfo{year}{2006}.
\newblock \bibinfo{title}{{A}n introduction to copulas}.
\newblock {S}pringer series in statistics, \bibinfo{publisher}{Springer},
  \bibinfo{address}{New York}.
\bibitem[{Nikoloulopoulos et~al.(2012)Nikoloulopoulos, Joe and
  Li}]{Nikoloulopoulos2012}
\bibinfo{author}{Nikoloulopoulos, A.K.}, \bibinfo{author}{Joe, H.},
  \bibinfo{author}{Li, H.}, \bibinfo{year}{2012}.
\newblock \bibinfo{title}{{V}ine copulas with asymmetric tail dependence and
  applications to financial return data}.
\newblock \bibinfo{journal}{Computational Statistics {\&} Data Analysis}
  \bibinfo{volume}{56}, \bibinfo{pages}{3659--3673}.
\bibitem[{Patton(2006)}]{Patton2006b}
\bibinfo{author}{Patton, A.J.}, \bibinfo{year}{2006}.
\newblock \bibinfo{title}{{M}odelling {A}symmetric {E}xchange {R}ate
  {D}ependence}.
\newblock \bibinfo{journal}{International Economic Review}
  \bibinfo{volume}{47}, \bibinfo{pages}{527--556}.
\bibitem[{Song et~al.(2009)Song, Li and Yuan}]{Song2009}
\bibinfo{author}{Song, P.X.K.}, \bibinfo{author}{Li, M.},
  \bibinfo{author}{Yuan, Y.}, \bibinfo{year}{2009}.
\newblock \bibinfo{title}{{J}oint {R}egression {A}nalysis of {C}orrelated
  {D}ata {U}sing {G}aussian {C}opulas}.
\newblock \bibinfo{journal}{Biometrics} \bibinfo{volume}{65},
  \bibinfo{pages}{60--68}.
\bibitem[{{Spanhel} and {Kurz}(2015)}]{Spanhel2015}
\bibinfo{author}{{Spanhel}, F.}, \bibinfo{author}{{Kurz}, M.S.},
  \bibinfo{year}{2015}.
\newblock \bibinfo{title}{{Simplified vine copula models: Approximations based
  on the simplifying assumption}}.
\newblock \bibinfo{journal}{ArXiv e-prints}
  \href{http://arxiv.org/abs/1510.06971}{\tt arXiv:1510.06971}.
\bibitem[{St{\"o}ber et~al.(2013)St{\"o}ber, Joe and Czado}]{Stober2013b}
\bibinfo{author}{St{\"o}ber, J.}, \bibinfo{author}{Joe, H.},
  \bibinfo{author}{Czado, C.}, \bibinfo{year}{2013}.
\newblock \bibinfo{title}{{S}implified pair copula
  constructions---{L}imitations and extensions}.
\newblock \bibinfo{journal}{Journal of Multivariate Analysis}
  \bibinfo{volume}{119}, \bibinfo{pages}{101--118}.

\end{thebibliography}
